\documentclass[3p,times]{elsarticle}

\usepackage{ecrc}

\usepackage{epstopdf}

\volume{00}

\firstpage{1}

\journalname{Nuclear Physics A}

\runauth{Barbara Betz et al.}

\jid{nupha}

\jnltitlelogo{Nuclear Physics A}

\usepackage{graphicx}
\usepackage{amsmath,amssymb}

\begin{document}

\begin{frontmatter}



\title{Azimuthal Jet Tomography at RHIC and LHC}

\author[label1]{Barbara Betz}
\author[label2]{Miklos Gyulassy}

\address[label1]{Institute for Theoretical Physics, Johann Wolfgang 
Goethe-University, 60438 Frankfurt am Main, Germany}
\address[label2]{Department of Physics, Columbia University, New York, 
10027, USA}

\begin{abstract}
Results based on a generic jet-energy loss model that interpolates between running
coupling pQCD-based and AdS/CFT-inspired holographic prescriptions are 
compared to recent data on the high-$p_T$ pion nuclear modification factor and 
the high-$p_T$ elliptic flow in nuclear collisions at RHIC and LHC. The 
jet-energy loss model is coupled to various (2+1)d (viscous hydrodynamic) fields. 
The impact of energy-loss fluctuations is discussed. While a previously proposed 
AdS/CFT jet-energy loss model with a temperature-independent jet-medium coupling 
is shown to be inconsistent with the LHC data, we find a rather broad class of jet-energy 
independent energy-loss models $dE/dx= \kappa(T) x^z T^{2+z}$ that can account for 
the current data with different temperature-dependent jet-medium couplings 
$\kappa(T)$ and path-length dependence exponents of $0\le z \le 2$.
\end{abstract}


\end{frontmatter}


\section{Introduction}

Jet-quenching observables are considered as significant probes of the 
medium evolution during a heavy-ion collision. Below, we investigate
a generic jet-energy loss model coupled to different temperature fields
of the quark-gluon plasma (QGP) that interpolates between running coupling 
pQCD-based models and AdS/CFT-inspired holographic prescriptions. 
The results are compared to recent data on the nuclear modification factor 
$R_{AA}$ and the high-$p_T$ elliptic flow \cite{data1,data2}, examining the 
dependence on the transverse momentum, the azimuthal angle, the centrality, 
and the collision energy to study the jet-medium coupling
$\kappa$ and the impact of the jet path-length dependence. The latter one
is usually considered to differentiate pQCD-based and AdS/CFT-inspired 
prescriptions with a linear and a squared path-length dependence, respectively.

The generic energy loss model studied is parametrized as 
\begin{eqnarray}
\frac{dE}{dx}=\frac{dE}{d\tau}(\vec{x}_0,\phi,\tau)= 
-\kappa(T)  E^a(\tau) \, \tau^{z} \, T^{c=2+z-a} \, \zeta_q
\;,
\label{Eq1}
\end{eqnarray}
with the jet-energy dependence, the path-length dependence, and the temperature 
dependence being characterized by the exponents $(a,z,c)$. The jet-medium 
coupling, $\kappa(T)$, may depend non-monotonically on the local temperature field,
where $\kappa(T)=C_r\kappa^\prime(T)$ for quark $C_r=1$ 
and gluon $C_r=\frac{9}{4}$ jets.
$T=T[\vec{x}(\tau)=\vec{x}_0+ (\tau-\tau_0) \hat{n}(\phi),\tau]$
describes the local temperature along the jet path at time $\tau$ for a jet
initially produced time $\tau_0$. 

The jets are distributed according to a transverse initial profile specified by 
the bulk QGP flow fields given by three variants of transverse plus Bjorken 
(2+1)d expansion \cite{background}: (1) VISH2+1, (2) viscous RL hydro, and (3) a 
$v_\perp=0.6$ blast wave flow assuming radial dilation of the initial transverse 
profile.

\begin{figure}[t]
\begin{center}
\includegraphics*[width=15cm]{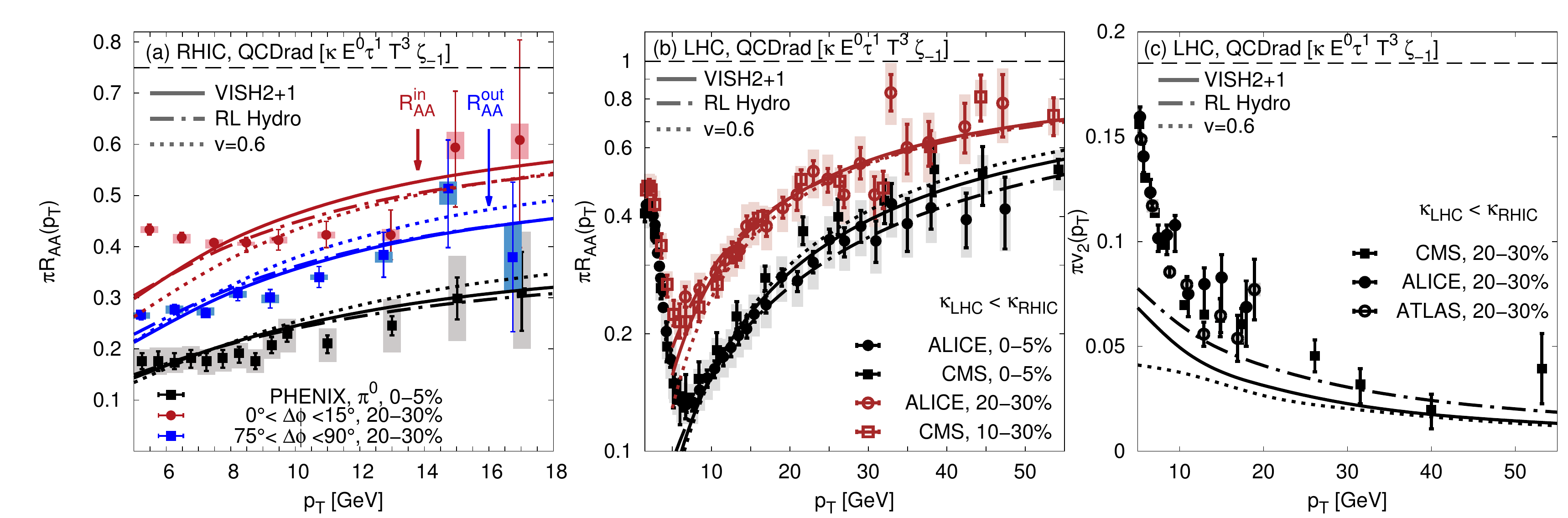}
\caption{The nuclear modification factor in- and out-of-plane at RHIC (left panel)
as well as the nuclear modification factor (middle panel) and the high-$p_T$ elliptic flow
(right panel) at the LHC for a pQCD-based energy loss 
$dE/d\tau=\kappa E^0 \tau^1 T^3 \zeta_{-1}$ without jet-energy loss fluctuations for 
various backgrounds \cite{background} compared to the measured data \cite{data1,data2}.
\vspace*{-0.4cm}}
\label{Fig01}
\end{center}
\end{figure}

We include skewed jet-energy loss fluctuations about the path-averaged mean by
using a scaling factor $\zeta_q$ describing the distributions according to 
$f_q(\zeta_q)= \frac{(1 + q)}{(q+2)^{1+q}} (q + 2- \zeta_q)^q $.
This class of skewed distributions is controlled by a parameter $q>-1$ that 
interpolates between non-fluctuating ($q=-1$, $\zeta_{-1}=1$), uniform Dirac 
distributions and distributions increasingly skewed towards small $\zeta_q < 1$ 
for $q>-1$, similar to pQCD based models \cite{us}.

The above jet-energy loss prescription allows for an easy comparison of 
perturbative QCD (pQCD) based models with exponents $(0,1,3,q)$ \cite{us} and conformal 
AdS holography models with non-linear path length $(0,2,4,q)$ \cite{us,AdS}, 
as well as phenomenological models based on a $\kappa(T)$ and $\kappa(\phi)$ \cite{us} as 
discussed below.

\section{Results}

Fig.\ \ref{Fig01} shows the results for a pQCD-based model $(0,1,3,-1)$
without energy-loss fluctuations for the three different transverse backgrounds
mentioned above both at RHIC (left panel) and at LHC energies (middle and right 
panel). Please note that the $R_{AA}$ in- and out-of-plane 
[$R_{AA}^{\rm in/out}=R_{AA}(1\pm 2v_2)$] allows for a simultaneous prescription
of the measured nuclear modification factor {\it and} the high-$p_T$ elliptic
flow. Here, the elliptic flow is given by the size of the gap between the nuclear
modification factor in- and out-of-plane.

\begin{figure}[b]
\begin{center}
\includegraphics*[width=15cm]{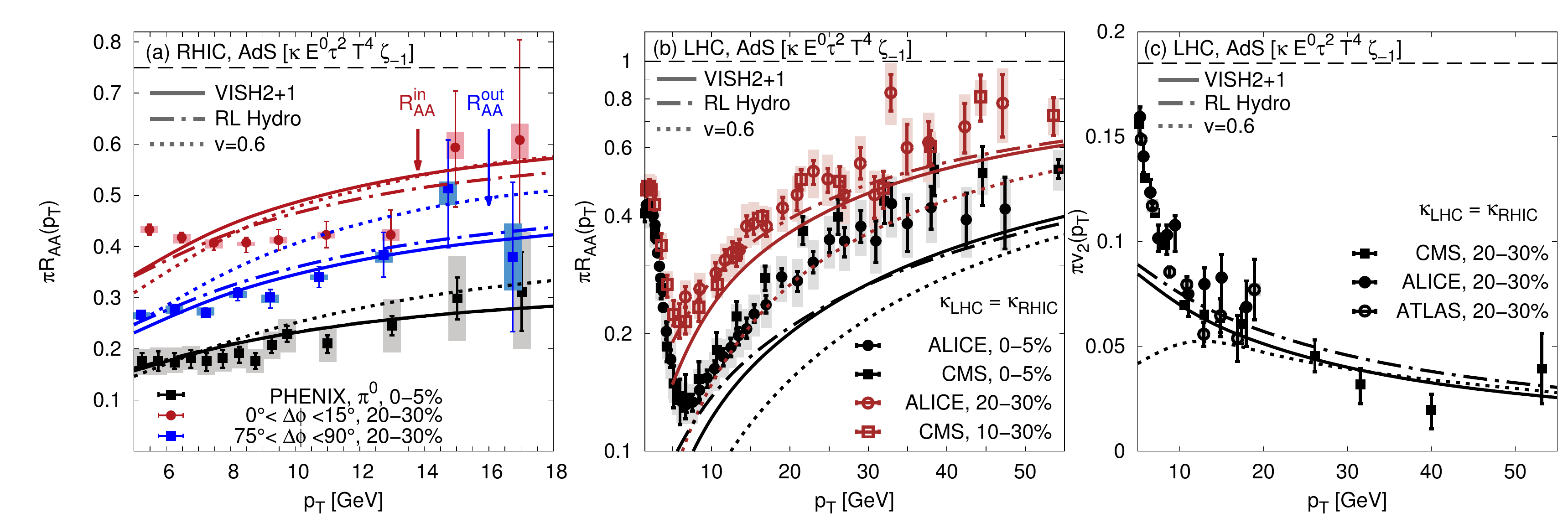}
\caption{The nuclear modification factor in- and out-of-plane at RHIC (left panel)
as well as the nuclear modification factor (middle panel) and the high-$p_T$ elliptic flow
(right panel) at the LHC for an AdS/CFT-inspired energy loss 
$dE/d\tau=\kappa E^0 \tau^2 T^4 \zeta_{-1}$ without jet-energy loss fluctuations for 
various backgrounds \cite{background} compared to the measured data \cite{data1,data2}.
\vspace*{-0.4cm}}
\label{Fig02}
\end{center}
\end{figure}

Fig.\ \ref{Fig01} clearly reveals that the results for both hydrodynamic prescriptions
reproduce the measured data within the uncertainties of the bulk space-time
evolution, given by the initial conditions, the viscosity ($\eta/s$), the
intial time $\tau_0$, the freeze-out time $T_f$, etc.\ considered. For this
pQCD-based model we account for a running coupling \cite{RunCoup,Xu} explaining the 
``surprising transparency''of the QGP at the LHC. 

Fig.\ \ref{Fig02} demonstrates that the {\it conformal} AdS/CFT-inspired 
prescription $(0,2,4,-1)$ without energy-loss fluctuations does certainly serve as
a description of the measured data at RHIC energies but does not reproduce 
the observed transparency at LHC by overpredicting the jet-energy loss
as shown in the middle panel of Fig.\ \ref{Fig02}. Please note that a conformal
theory does not have an additional scale, i.e.\ the coupling considered {\rm cannot}
run, in contrast to the pQCD scenario considered in Fig.\ \ref{Fig01}. If, however,
a {\it non-conformal} AdS/CFT prescription is introduced allowing for a 
reduction of the running-coupling constant (at about the same magnitude
as used for pQCD prescriptions), this {\it novel, non-conformal} ansatz will also
describe both the nuclear modification factor and the high-$p_T$ elliptic flow
measured at RHIC and at LHC considering a (viscous) hydrodynamic background, 
as was shown in Refs.\ \cite{us}.

Moreover, Fig.\ \ref{Fig02} reveals that the background described by the simple 
blast wave model can also be ruled out as it usually leads to an elliptic flow 
that is too small as compared to data \cite{us}.

For pQCD, a reduction of the effective jet-medium coupling with $\sqrt{s}$ is natural 
due to the vacuum running of the perturbative QCD coupling $\alpha_s(Q)$. 
However, even after this reduction is taken into account, there is a tendency
shown in Fig.\ \ref{Fig01} that the high-$p_T$ elliptic flow is underpredicted 
as compared to the measured data. This is in line with various pQCD-based
models (AMY, HT, ASW, Molnar, CUJET2.0) \cite{data1,Xu,Molnar} that are 
below the data. 

To overcome this ``high-$p_T$ $v_2$ problem'' of pQCD-based jet-energy loss 
prescriptions, Ref.\ \cite{Xu} suggested that there could well be an
additional running of the running coupling constant w.r.t.\ the temperature, 
$\alpha_{\rm eff}(Q,T)$, as suggested by Lattice QCD \cite{Kaczmarek:2004gv}
that might cause a modest ($10-15\%$) variation of the path-averaged coupling 
in non-central collisions with a coupling constant enhanced out-of-plane. 
To simulate this effect, we study an azimuthal dependence of the jet-medium coupling by 
$\kappa(\phi) = \kappa\cdot(1+\vert\sin(\phi)\vert\cdot X)$, 
where $X$ is a given value in percentage. 

\begin{figure}
\begin{center}
\includegraphics*[width=15cm]{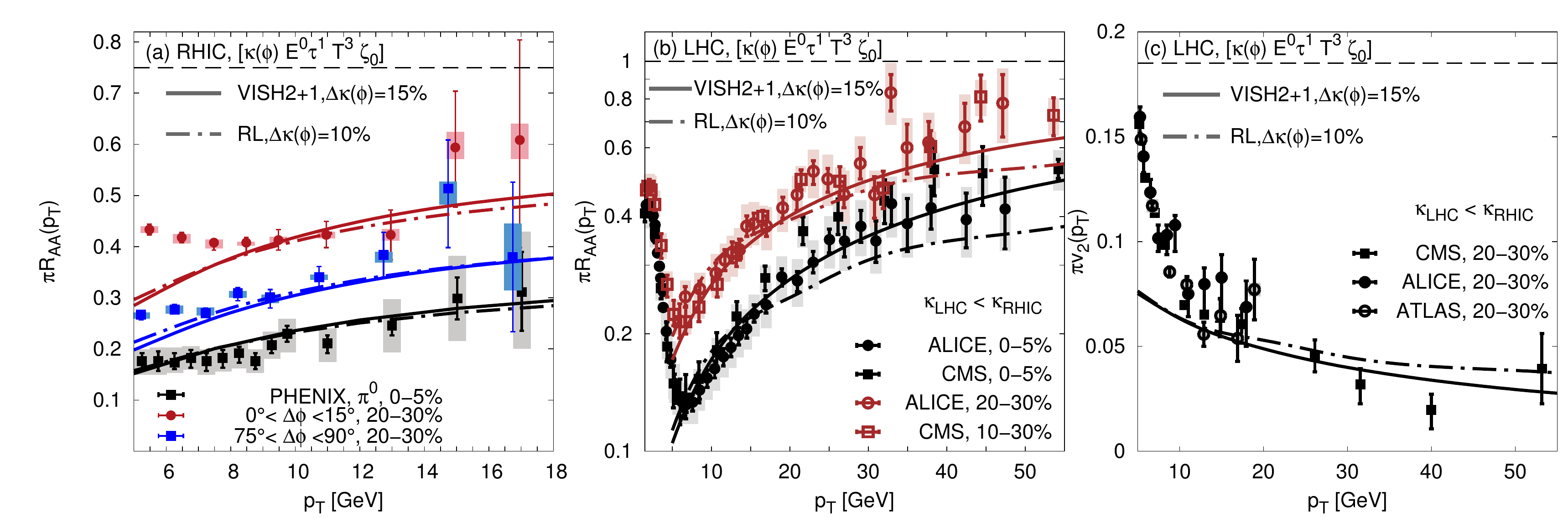}
\caption{The nuclear modification factor in- and out-of-plane at RHIC (left panel)
as well as the nuclear modification factor (middle panel) and the high-$p_T$ elliptic flow
(right panel) at the LHC for a pQCD-based energy loss 
$dE/d\tau=\kappa(\phi) E^0 \tau^1 T^3 \zeta_{0}$ with an azimuthal variation of the 
jet-medium coupling for two different hydrodynamic backgrounds 
\cite{background} compared to the measured data \cite{data1,data2}.
\vspace*{-0.8cm}}
\label{Fig03}
\end{center}
\end{figure}

Fig.\ \ref{Fig03} demonstrates that indeed a small azimuthal variation of $10-15\%$ of the 
jet-medium coupling is sufficient to account for the high-$p_T$ $v_2$ problem at 
the LHC while it simultaneously allows for a decent description of the nuclear modification
factor and the $R_{AA}^{\rm in/out}$ at RHIC. Here we considered again the pQCD-
inspired prescription $(0,1,3,0)$ including energy-loss fluctuations. As 
shown in Ref.\ \cite{us} (and Fig.\ \ref{Fig04}), the influence of the jet-energy loss fluctuations 
themselves is significant for the yield of the nuclear modification factor but
not for the high-$p_T$ elliptic flow.

The fact that a moderate azimuthal dependence of the jet-medium coupling
with a coupling enhanced out-of-plane can describe the measured data, 
in general supports a jet-medium coupling enhanced for lower temperatures 
\cite{Liao:2008dk} as a jet traversing out-of-plane will propagate longer 
through a comparably cooler medium. 

In Fig.\ \ref{Fig04} we exemplary consider an exponentially falling ansatz 
for the jet-medium coupling, $\kappa(T) = \kappa_1 e^{-b(T-T_1)}$, assuming 
that the coupling is zero below a certain temperature $T_1$, representing the freeze-out. 
At this $T_1$ the coupling peaks at a value of $\kappa_1$ and then falls off for 
larger temperatures to a value of $1/e$ at a temperature $T_e$ \cite{us}. Please note
that this exponential drop of the jet-medium coupling implies an {\em effective} 
reduction of the jet-medium coupling at the LHC as compared to RHIC. 

The results shown in Fig.\ \ref{Fig04} demonstrate that this exponential ansatz
for the jet-medium coupling also describes the measured data within the present
error bars even though the high-$p_T$ elliptic flow is again rather low. 
Surprisingly enough, the values of $T_1=160$~MeV and $T_e=200$~MeV indicate 
that the high-temperature medium is basically transparent \cite{us,RunCoup,Xu}. 

\section{Conclusions}

\begin{figure}
\begin{center}
\includegraphics*[width=15cm]{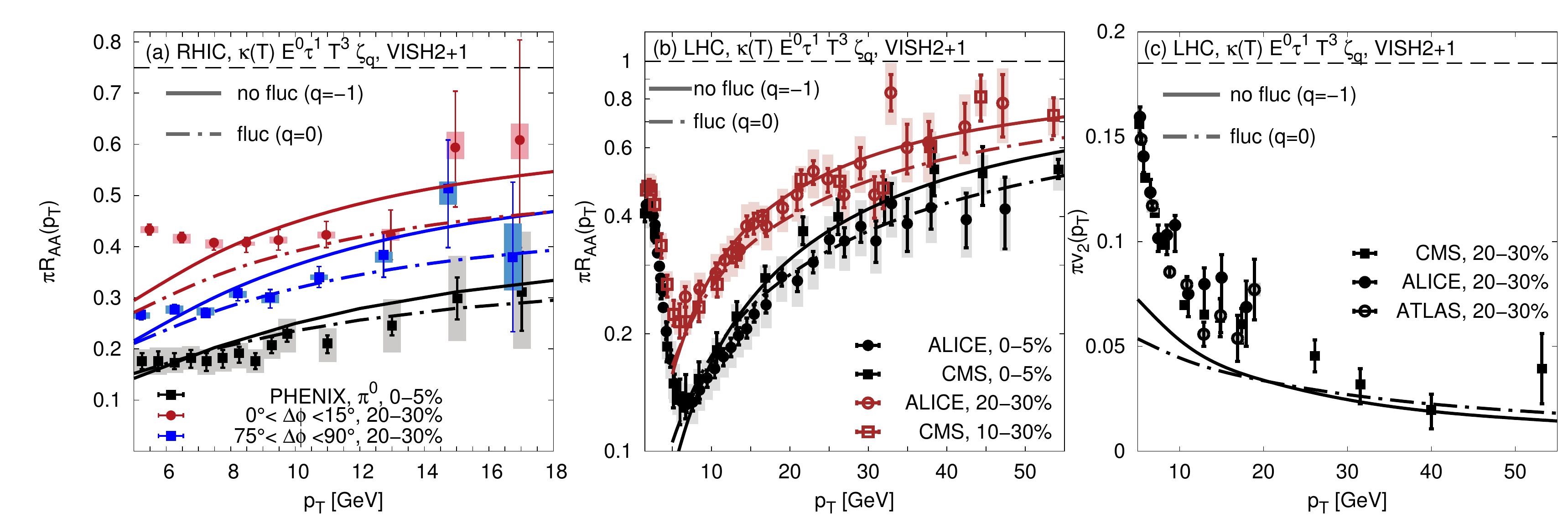}
\caption{The nuclear modification factor in- and out-of-plane at RHIC (left panel)
as well as the nuclear modification factor (middle panel) and the high-$p_T$ elliptic flow
(right panel) at the LHC for a pQCD-based energy loss 
$dE/d\tau=\kappa(T) E^0 \tau^1 T^3 \zeta_{q}$ for a temperature-dependent jet-medium 
coupling with ($q=0$) and without ($q=-1$) flucutations and a hydrodynamic background
\cite{background} compared to the measured data \cite{data1,data2}.
\vspace*{-0.8cm}}
\label{Fig04}
\end{center}
\end{figure}

We compared the results of a generic jet-energy loss model, 
$dE/d\tau = -\kappa(T) E^a \tau^{z} T^{c=2+z-a} \zeta_q$, that
interpolates between pQCD-based $(a=0,z=1,c=3)$ and AdS/CFT-inspired 
$(a=0,z=2,c=4)$ prescriptions to recent data \cite{data1,data2} on the 
nuclear modification factor and the high-$p_T$ elliptic flow both at 
RHIC and at LHC energies. The model discussed includes possible jet-energy loss 
fluctuations via the multiplicative factor $\zeta_q$. We found that the 
running coupling energy-loss models motivated by pQCD appears to be favored, 
while the AdS/CFT-inspired ansatz excluding by first principles the running 
of the coupling constant $\kappa$ cannot reproduce the measured data.
We showed that the running of the coupling constant either with collision energy
(Fig.\ \ref{Fig01}) or with temperature (Fig.\ \ref{Fig04}) reproduces
the measured data within the present error bars considering (viscous)
hydrodynamic background fields \cite{background}. However, these prescriptions 
do not describe the high-$p_T$ elliptic flow with best accuracy which
can only be reachted (Fig.\ \ref{Fig03}) once an additional azimuthal
dependence of the jet-medium coupling \cite{us,Xu} is considered with
a coupling enhanced out-of-plane.

\section{Acknowledgement}
This work was supported in part through the Helmholtz International Centre for FAIR 
within the framework of the LOEWE program (Landesoffensive zur Entwicklung 
Wissenschaftlich-\"Okonomischer Exzellenz) launched by the State of Hesse, 
the US-DOE Nuclear Science Grant No.\ DE-AC02-05CH11231 within the framework 
of the JET Topical Collaboration, and the US-DOE Nuclear Science Grant No.\
DE-FG02-93ER40764.

\end{document}